\documentclass[aps,prl,prabib,twocolumn,
showpacs,preprintnumbers,amsmath,amssymb,floatfix,aps]{revtex4}

\usepackage{graphicx}

\begin{document}

\title{Phase separation in a polarized Fermi gas at zero temperature}
\author{S. Pilati and S. Giorgini}
\affiliation{Dipartimento di Fisica, Universit\`a di Trento
and CNR-INFM BEC Center, I-38050 Povo, Trento, Italy}

\begin{abstract} 
We investigate the phase diagram of asymmetric two-component Fermi gases at zero temperature as a function of polarization and interaction strength. The equations of state of the uniform superfluid and normal phase are determined using quantum Monte Carlo simulations. We find three different mixed states, where the superfluid and the normal phase coexist in equilibrium, corresponding to phase separation between: (a) the polarized superfluid and the fully polarized normal gas, (b) the polarized superfluid and the partially polarized normal gas and (c) the unpolarized superfluid and the partially polarized normal gas. 
\end{abstract}

\maketitle 

The study of a two-component Fermi gas with imbalanced populations is an active area of research in the field of ultracold atoms~\cite{RMP}. Recent experiments, carried out on harmonically trapped configurations, investigate superfluidity and Bose-Einstein condensation (BEC) of fermionic pairs in these systems by varying the strength of interactions, the temperature of the gas and the degree of polarization~\cite{MIT}. Evidences of phase separation between a superfluid core and a normal external shell are reported for systems close to the unitary limit~\cite{Partridge06} and on both sides of the Feshbach resonance~\cite{Shin07}. On the theoretical side, the phase diagram of a polarized Fermi gas with tunable interactions is the subject of a number of studies both at zero~\cite{PhaseDiagram1,PhaseDiagram2} and at finite temperature~\cite{Parish07}. These studies, which refer to uniform systems and are based on a mean-field approach, predict the existence of a normal phase for large enough polarization on the BCS side of the crossover and of a polarized superfluid phase on the BEC side, separated by a region where the two phases coexist in equilibrium. However, if interactions are not weak, the mean-field theory fails to describe correctly the nature of the phase separated state and to provide a reliable estimate of the critical value of polarization where phase separation occurs. The key ingredient, that is missing in the mean-field description, is the proper account of interaction effects in the normal phase~\cite{PhaseSeparation2}.

In this Letter we carry out a quantitative study of the phase diagram at zero temperature along the BCS-BEC crossover using fixed-node diffusion Monte Carlo (FN-DMC) simulations, which have proven very accurate in the investigation of strongly correlated ultracold Fermi gases~\cite{RMP}. We determine the equation of state of the homogeneous normal and superfluid phase as a function of interaction stregth and population imbalance. From the phase equilibrium conditions we identify three different phase separated states corresponding to: (a) the polarized superfluid coexisting with the fully polarized normal gas, (b) the polarized superfluid coexisting with the partially polarized normal gas and (c) the unpolarized superfluid coexisting with the partially polarized normal gas. The state (a) and (c) are respectively separated from the homogeneous superfluid and normal phase by a first order phase transition, while state (a) and state (b) as well as state (b) and state (c) are separated by second order phase transitions. 

We consider a uniform system in a volume $V$ with a total number of particles $N=N_\uparrow+N_\downarrow$. The number $N_\uparrow$ of spin-up particles is kept fixed, corresponding to the average density $n_\uparrow=N_\uparrow/V$. The interaction strength is parametrized by the inverse product $1/k_{F\uparrow}a$ of the $s$-wave scattering length $a$ and the Fermi wavevector of the spin-up particles $k_{F\uparrow}=(6\pi^2n_\uparrow)^{1/3}$. The number $N_\downarrow$ of spin-down particles is instead a variable of the system determining the polarization parameter $P=\frac{N_\uparrow-N_\downarrow}{N_\uparrow+N_\downarrow}$, that is assumed positive within the bounds $0\leq P\leq 1$. The relevant energy scale is fixed by the Fermi energy of the spin-up particles $E_{F\uparrow}=\hbar^2k_{F\uparrow}^2/2m$, where $m$ is the particle mass of both spin components.

First, we discuss the equation of state of the four homogeneous phases considered in the present study: unpolarized and polarized superfluid and fully and partially polarized normal gas. Then, we analyze the equilibrium conditions between these phases to map out the phase diagram along the BCS-BEC crossover. We do not consider more exotic superfluid phases such as the Fulde-Ferrel and Larkin-Ovchinnikov (FFLO) state which is expected to occur on the BCS side of the resonance for small polarizations~\cite{Combescot07_1}. 

{\it i) Fully polarized normal gas} (N$_\text{FP}$).

In this phase $N_\downarrow=0$. Since $p$-wave collisions can be neglected, the gas is well described by the non-interacting model with the equation of state
\begin{equation}
E_{\text{N}_\text{FP}}=\frac{3}{5}N_\uparrow E_{F\uparrow} \;.
\label{EOS1}
\end{equation}

{\it ii) Unpolarized superfluid gas} (SF$_0$).

\begin{figure}
\begin{center}
\includegraphics[width=7cm]{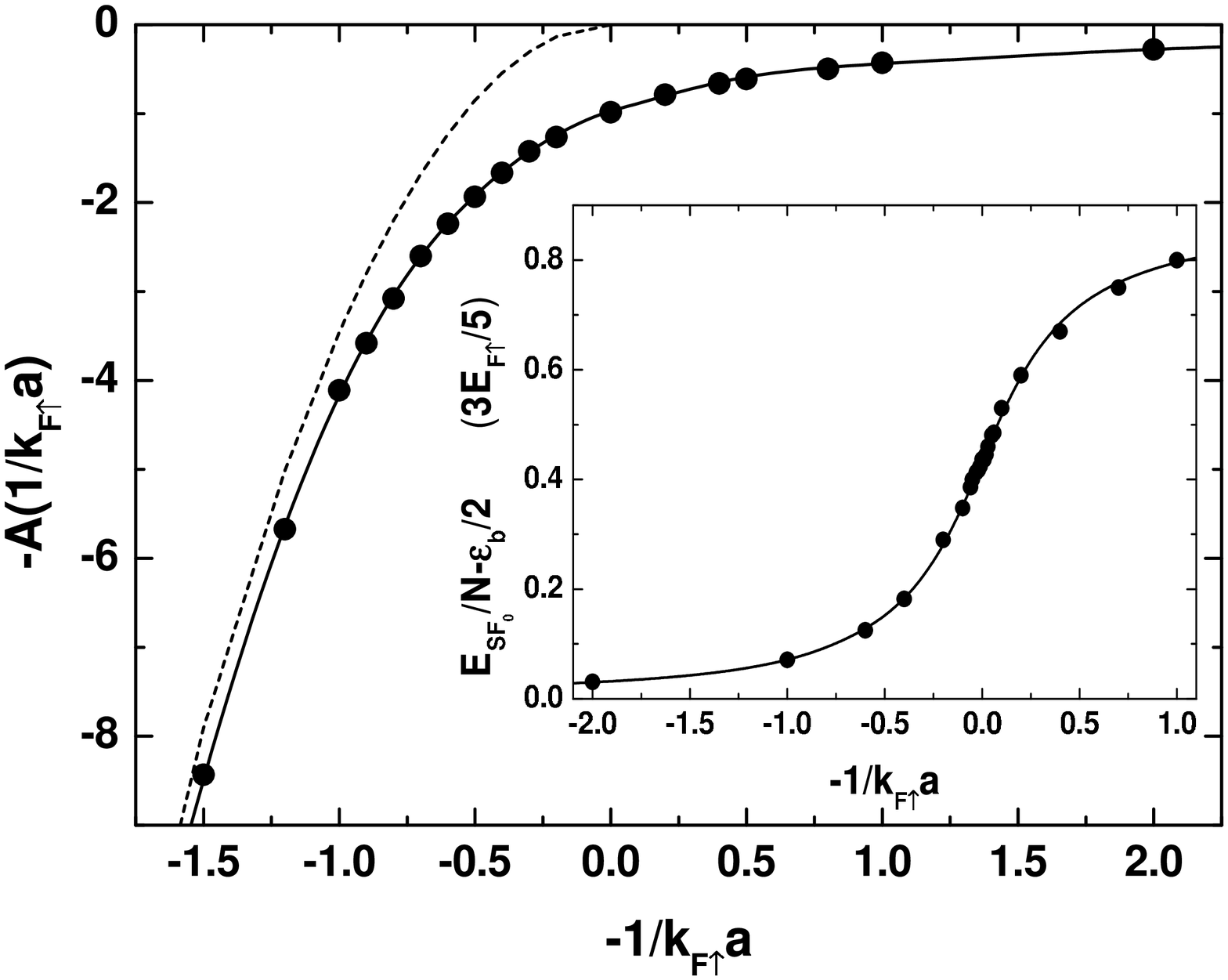}
\caption{Binding energy $A$ of a single spin-down impurity in a Fermi sea of spin-up particles. The dashed line is the molecular binding energy $\epsilon_b$ for our short-range square well potential (with $2n_\uparrow R_0^3=10^{-6}$ as in Ref.~\cite{Lobo06}). In the inset we show the equation of state of the unpolarized superfluid SF$_0$. The solid lines correspond to best fits to the FN-DMC results.}
\label{fig1}
\end{center}
\end{figure}

In this phase $N_\uparrow=N_\downarrow=N/2$. The corresponding ground-state energy can be cast in the following form
\begin{equation}
E_{\text{SF}_0}=\frac{N}{2}\epsilon_b\Theta(1/k_{F\uparrow}a) + \left(\frac{3}{5}N_\uparrow E_{F\uparrow}\right) 
2G(1/k_{F\uparrow}a) \;.
\label{EOS2}
\end{equation}
The first term, proportional to the number of pairs $N/2$, corresponds to the contribution of molecules with binding energy $\epsilon_b$. For a zero-range interatomic potential this energy takes the familiar expression $\epsilon_b=-\hbar^2/ma^2$. The two-body term is a convenient parametrization of the equation of state only on the BEC side of the crossover ($a>0$) where dimers are formed in vacuum, as entailed in Eq.~(\ref{EOS2}) by the Heaviside function: $\Theta(x)=1$ if $x>0$ and zero otherwise. The dimensionless function of the interaction strength $G(1/k_{F\uparrow}a)$ contains instead the many-body contributions to the equation of state. This function has been calculated in Ref.~\cite{Astrakharchik04} using the FN-DMC method and the results are shown in the inset of Fig.~\ref{fig1}. In particular, in the BEC regime ($1/k_{F\uparrow}a\gg 1$) the function $G$ takes the form $G=5k_{F\uparrow}a_{dd}/18\pi[1+128(k_{F\uparrow}a_{dd}/\pi)^{3/2}/15\sqrt{6}]$ with $a_{dd}=0.60a$~\cite{Petrov04} and corresponds to the mean-field and first beyond mean-field contributions to the equation of state of composite bosons with mass $2m$ and density $n_\uparrow$ interacting with a dimer-dimer scattering length $a_{dd}$.

{\it iii) Partially polarized normal gas} (N$_\text{PP}$).

\begin{figure}
\begin{center}
\includegraphics[width=7cm]{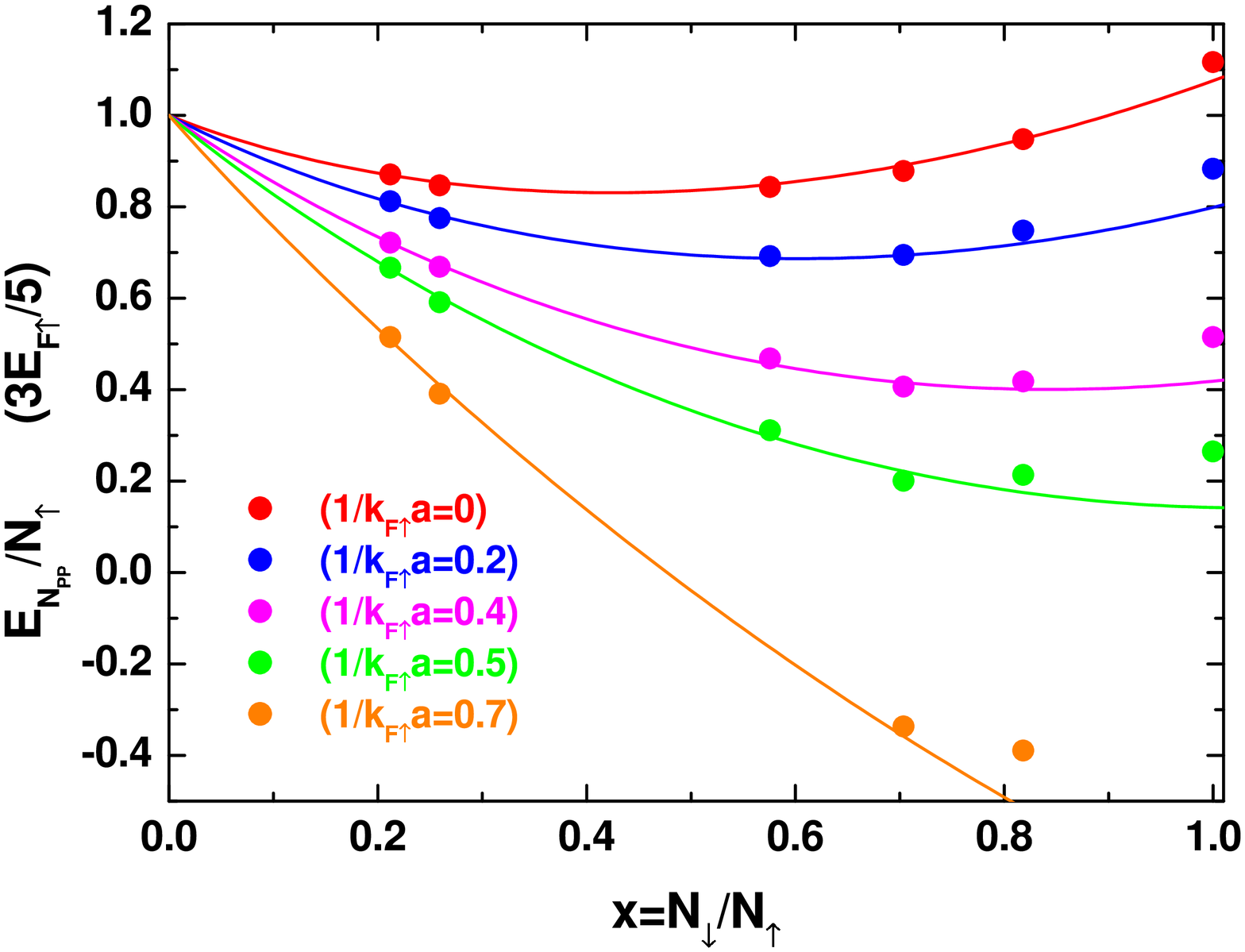}
\caption{(color online). Equation of state of the normal partially polarized phase N$_\text{PP}$ as a function of the concentration $x$ for different values of the interaction strength. The solid lines correspond to best fits of the energy functional (\ref{EOS3}) with the values of $A$ and $m^\ast$ obtained from the single-impurity calculations.}
\label{fig2}
\end{center}
\end{figure}

This phase is characterized by the concentration $x=N_\downarrow/N_\uparrow$ of the minority spin-down particles. At small concentrations ($x\ll 1$) the dependence on $x$ of the ground-state energy can be written in the form of the Landau-Pomeranchuk Hamiltonian of weakly interacting quasiparticles~\cite{Lobo06}  
\begin{equation}
E_{\text{N}_\text{PP}}=\frac{3}{5}N_\uparrow E_{F\uparrow} \left(1-Ax+\frac{m}{m^\ast}x^{5/3}+Fx^2\right)\;,
\label{EOS3}
\end{equation}
where $A$, $m^\ast/m$ and $F$ depend on $1/k_{F\uparrow}a$. The quantity $A$ is the binding energy of a spin-down quasiparticle in the Fermi sea of spin-up particles and $m^\ast$ is its effective mass. The term $F$ accounts instead for the coupling between quasiparticles. As already pointed out, the mean-field approach completely neglects interactions in this phase resulting in an energy functional given by Eq.~(\ref{EOS3}) with $A=F=0$ and $m^\ast/m=1$. We determine $A$ and $m^\ast$ as a function of the interaction strength from quantum Monte Carlo simulations. The binding energy is obtained from the ground-state energy of the system with one spin-down impurity in a Fermi sea of spin-up particles and the effective mass from the curvature of the excitation energy if the impurity carries a small momentum~\cite{Lobo06}. The results for $A$ are shown in Fig.~\ref{fig1}. They are in excellent agreement with the values recently obtained using exact diagrammatic Monte Carlo methods~\cite{ProkofevSvistunov07} and, quite remarkably, also with the results of a simple variational approach based on a single particle-hole wavefunction~\cite{Combescot07}. The values we obtain for the effective mass $m^\ast$ are instead slightly smaller than the diagrammatic Monte Carlo results of Ref.~\cite{ProkofevSvistunov07}. This might be due to a non optimal choice of the nodal surface of the excited state at finite momentum, the FN-DMC method provides in fact only an upper bound for the energy unless the nodes of the many-body wavefunction are known exactly, and to finite size effects in the analysis of the low-momentum spectrum~\cite{note0}. We also perform FN-DMC calculations at finite concentration $x$ for various values of the interaction parameter $1/k_{F\uparrow}a$ using the Jastrow-Slater wavefunction described in Ref.~\cite{Lobo06}. The results are presented in Fig.~\ref{fig2}. By fitting the functional form (\ref{EOS3}) to these results with $A$ and $m^\ast$ obtained from the single-impurity calculations, we determine the interaction parameter $F$ and its dependence as a function of $1/k_{F\uparrow}a$. At unitarity we find: $A(1/k_{F\uparrow}a=0)=0.99(1)$, $m^\ast(1/k_{F\uparrow}a=0)/m=1.09(2)$ and $F(1/k_{F\uparrow}a=0)=0.14$, in agreement with the findings of Ref.~\cite{Lobo06}. We notice that the N$_\text{PP}$ phase reduces to the N$_\text{FP}$ one if $x=0$.

\begin{figure}
\begin{center}
\includegraphics[width=7cm]{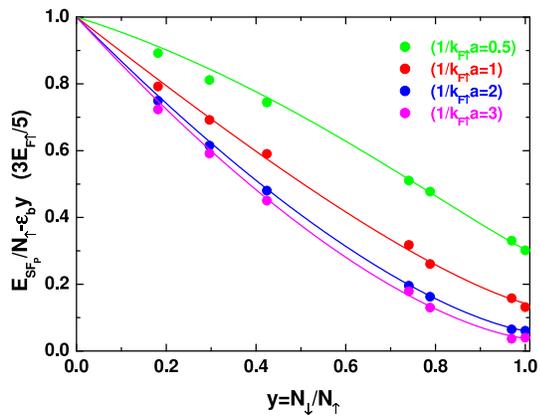}
\caption{(color online). Equation of state of the superfluid polarized phase SF$_\text{P}$ as a function of the concentration $y$ for different values of the interaction strength. The solid lines correspond to the energy functional (\ref{EOS4}).}
\label{fig3}
\end{center}
\end{figure}

{\it iv) Polarized superfluid gas} (SF$_\text{P}$).

This phase is characterized by a number $N_P=N_\downarrow$ of pairs and a number $N_A=N_\uparrow-N_\downarrow$ of unpaired atoms, such that $N=2N_P+N_A$. We denote the concentration of the minority atoms by $y=N_\downarrow/N_\uparrow$. In the deep BEC regime the SF$_\text{P}$ phase corresponds to a miscible mixture of $N_P$ bosons and $N_A$ fermions~\cite{Viverit00,BoseFermi}. The interaction between bosons and fermions is repulsive and is fixed by the atom-dimer scattering length $a_{ad}=1.18a$~\cite{SkorniakovTerMartirosian57}. In this regime we write the equation of state in the form
\begin{eqnarray}
E_{\text{SF}_\text{P}}&=&E_{\text{SF}_0}(N_P) 
\label{EOS4} \\
&+&\frac{3}{5}N_\uparrow E_{F\uparrow} \left[(1-y)^{5/3} + \frac{5k_{F\uparrow}a_{ad}}{3\pi}y(1-y) \right]\;,
\nonumber
\end{eqnarray}
where $E_{\text{SF}_0}(N_P)=3/5N_\uparrow E_{F\uparrow}y^{5/3} 2 G(1/k_{F\uparrow}ay^{1/3})+\epsilon_bN_P$ is the energy of the $2N_P$ paired atoms and the other terms in Eq.(\ref{EOS4}) correspond to the kinetic energy of the unpaired atoms and to the interaction energy between atoms and dimers treated at the mean-field level. We carry out FN-DMC simulations of the SF$_\text{P}$ phase for various values of the interaction strength. The nodal surface is modeled using a BCS plus unpaired particles wavefunction written in the form of a determinant as in Ref.~\cite{CarlsonReddy05}. The results are shown in Fig.~\ref{fig3} together with the energy functional (\ref{EOS4}). The agreement is remarkable down to quite small values of the interaction parameter $1/k_{F\uparrow}a\geq0.5$\cite{note1}. Furthermore, we notice that the SF$_\text{P}$ phase reduces to the SF$_0$ one if $y=1$.

We are now in a position to study the coexistence between the superfluid and normal phases introduced above. One requires the equilibrium of pressures between the superfluid $p_S=-\partial E_S/\partial V_S$ and the normal $p_N=-\partial E_N/\partial V_N$ state and the equilibrium of chemical potentials. In the normal phase there are two chemical potentials for the N$_\text{PP}$ state: $\mu_{N\uparrow(\downarrow)}=\partial E_N/\partial N_{\uparrow(\downarrow)}$, which reduce to only $\mu_{N\uparrow}$ for the N$_\text{FP}$ state. Similarly in the SF$_\text{P}$ phase one can vary both the number of pairs $\mu_{SP}=\partial E_S/\partial N_P$ and the number of unpaired atoms $\mu_{SA}=\partial E_S/\partial N_A$, while in the SF$_0$ phase only the chemical potential of pairs $\mu_{SP}$ with $N_P=N/2$ is relevant.

\begin{figure}
\begin{center}
\includegraphics[width=7.5cm]{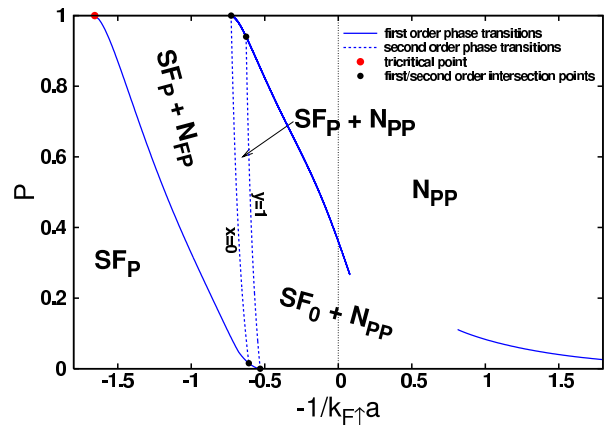}
\caption{(color online). Phase diagram as a function of polarization and interaction strength. In terms of the Fermi wavevector $k_F=(3\pi^2n)^{1/3}$ fixed by the total density $n=N/V$ one has: $1/k_Fa=1/k_{F\uparrow}a$ at $P=0$ and $1/k_Fa=2^{1/3}/k_{F\uparrow}a$ at $P=1$. On the BCS side of the resonance our determination of the critical polarization is not reliable. For $-1/k_{F\uparrow}a\gtrsim 1$ we obtain $P_c$ using the BCS theory (see text).}
\label{fig4}
\end{center}
\end{figure}

{\it a) Phase separation between} SF$_\text{P}$ {\it and} N$_\text{FP}$. The equilibrium conditions are: $p_S=p_N$ and  $\mu_{SA}=\mu_{N\uparrow}$. For a given concentration $y$ of the spin-down atoms in the SF$_\text{P}$ phase, the two conditions determine the values of the densities of the spin-up component in the two coexisting phases. The ratio $P_c=\frac{1-y}{1+y}$ gives the critical polarization above which the system begins nucleating the normal phase to accommodate the excess polarization. By increasing $P$ above $P_c$ the equilibrium densities of the two phases, as well as the critical concentration $y$ of the SF$_\text{P}$ phase, do not change, instead, the volume fraction $V_N/V$ of the normal phase increases and eventually becomes the entire volume for $P=1$. The critical polarization line, corresponding to a first order phase transition, is shown in the phase diagram of Fig.~\ref{fig4}. At $P=1$ this line terminates at the tricritical point $1/k_{F\uparrow}a=1.7$~\cite{note2}. For larger values of $1/k_{F\uparrow}a$ the homogenous SF$_\text{P}$ phase exists up to $P=1$ and the superfluid to normal transition becomes second order. For a given concentration $y$ the SF$_\text{P}$-N$_\text{FP}$ state is stable provided $\mu_{SP}\leq\mu_{N\uparrow}+\mu_{N\downarrow}=E_{F\uparrow}(1-3A/5)$, {\it i.e.} until the process in which pairs break and spin-down particles start to populate the normal phase remains energetically unfavourable. The instability line, corresponding to $x=0$, marks a second order phase transition where the fully polarized evolves continously into the partially polarized normal phase. At $P=1$ this line terminates at the point $1/k_{F\uparrow}a=0.73$, for smaller values of $1/k_{F\uparrow}a$ a superfluid can not exist up to $P=1$. For small polarizations the second order transition line terminates at the point corresponding to $P_c=0.015$ and $1/k_{F\uparrow}a=0.61$. 

{\it b) Phase separation between} SF$_\text{P}$ {\it and} N$_\text{PP}$. In this case one has to fulfill three equilibrium conditions: $p_S=p_N$, $\mu_{SA}=\mu_{N\uparrow}$ and $\mu_{SP}=\mu_{N\uparrow}+\mu_{N\downarrow}$. It is worth pointing out that the SF$_\text{P}$-N$_\text{PP}$ phase separated state does not exist within the mean-field description, where either the normal state is fully polarized or the superfluid state is unpolarized. By approaching the unitary limit the SF$_\text{P}$-N$_\text{PP}$ state becomes unstable because the polarization of the superfluid is energetically unfavourable and the SF$_\text{P}$ phase evolves continously into the SF$_0$ phase. The instability line, corresponding to $y=1$, marks another second order phase transition. This line terminates at the following points: $P=0.94$ and $1/k_{F\uparrow}a=0.63$ and $P=0$ and $1/k_{F\uparrow}a=0.53$. In particular, the point at $P=0$ corresponds to the smallest value of $1/k_{F\uparrow}a$ below which the superfluid phase can not be polarized~\cite{note3}. The two second order transition lines at $x=0$ and $y=1$ are indeed very close and, given the uncertainty in the determination of the equation of state of the various phases, we can not exclude that they might coincide, corresponding to a single second order phase transition where the superfluid polarizes and the normal phase becomes fully polarized, or that they might have a reversed order producing a small region of SF$_0$-N$_\text{FP}$ mixed phase instead of the SF$_\text{P}$-N$_\text{PP}$ one. It is important to stress, however, that in the relevant region we can provide a reliable description of the equation of state of the superfluid (see Fig.~\ref{fig3}) and of the normal phase, where for small concentration $x$ only the term containing the binding energy $A$ is important. 

{\it c) Phase separation between} SF$_0$ {\it and} N$_\text{PP}$. One has to fulfill two conditions: $p_S=p_N$ and $\mu_{SP}=\mu_{N\uparrow}+\mu_{N\downarrow}$. The SF$_0$-N$_\text{PP}$ state is stable provided $\mu_{N\uparrow}\leq\mu_{SA}$, corresponding to the instability against polarization of the superfluid. The instabilty line coincides with the second order phase transition for $y=1$ obtained above. The critical polarization line $P_c=\frac{1-x}{1+x}$ marks instead the first order phase transition from the normal to the superfluid gas. At unitarity we find $P_c=0.39$~\cite{Lobo06} in contrast with the value $P_c=0.93$ predicted by mean-field theory~\cite{PhaseDiagram2}. Notice that the Landau-Pomeranchuk energy functional (\ref{EOS3}) does not provide a valid description of the N$_\text{PP}$ phase if the concentration $x$ becomes large. For this reason, on the BCS side of the resonance, our results for $P_c$ are limited to the region close to the unitary limit. In the deep BCS regime one can calculate the critical polarization by using the following energy functionals for the N$_\text{PP}$ and SF$_0$ phase respectively~\cite{Bedaque03,CarlsonReddy05}: $E_{\text{N}_\text{PP}}=3N_\uparrow E_{F\uparrow}/5(1-20k_{F\uparrow}|a|x/9\pi)$ and $E_{\text{SF}_0}=E_{\text{N}_\text{PP}}-3N_\uparrow\Delta_\text{gap}^2/4E_{F\uparrow}$, where $\Delta_\text{gap}=(2/e)^{7/3}E_{F\uparrow}e^{-\pi/2k_{F\uparrow}|a|}$ is the superfluid gap. From the equilibrium conditions one obtains $P_c=3/\sqrt{8}(2/e)^{7/3}e^{-\pi/2k_{F\uparrow}|a|}$, holding to leading order in $P$.

In conclusion, we have investigated the phase diagram of a Fermi gas at $T=0$ as a function of polarization and interaction strength. This analysis, carried out for uniform gases, is relevant also for systems in harmonic traps that can be studied by means of the local density approximation.

We acknowledge useful discussions with N. Prokof'ev, C. Lobo, A. Recati and F. Chevy. We also aknowledge support by the Ministero dell'Universit\`a e della Ricerca.


\begin{thebibliography}{99}
\bibitem{note0} For the largest system we simulate ($N_\uparrow=33$) the smallest wavevector compatible with periodic boundary conditions we can assign to the impurity is only $q=0.50k_{F\uparrow}$.
\bibitem{note1} A similar remarkable agreement with the picture of compact dimers has been found in Ref.~\cite{ProkofevSvistunov07} for the energy of a molecule in a Fermi sea.
\bibitem{note2} The position of the tricritical point predicted by mean-field theory is at $1/k_{F\uparrow}a=1.88$~\cite{Parish07}.
\bibitem{note3} The position of this point predicted by mean-field theory is at $1/k_{F\uparrow}a=1.01$~\cite{PhaseDiagram2}.
\end{thebibliography}
\end{document}